\documentclass[twocolumn,showpacs,preprintnumbers,amsmath,amssymb]{revtex4}

\usepackage{graphicx}
\usepackage{rotating}
\usepackage{dcolumn}
\usepackage{bm}

\begin{document}
\date{\today}
\title{Electron cooling and Debye-Waller effect in photoexcited bismuth 
}
\author{B.~Arnaud$^{1}$ and Y. Giret$^{2,3}$}
\affiliation{$^{1}$Institut de Physique de Rennes (IPR),
UMR UR1-CNRS 6251, Campus de Beaulieu - Bat 11 A, 35042 Rennes Cedex, France, EU}
\affiliation{$^{2}$Institute of Scientific and Industrial Research, Osaka University,
8-1 Mihogaoka, Ibaraki, Osaka 567-0047, Japan}
\affiliation{$^{3}$Department of Physics and Astronomy, University College London, 
Gower Street, London WC1E 6BT, United Kingdom}

\date{\today}

\begin{abstract}
By means of first principles calculations, we computed the effective electron-phonon coupling
constant $G_0$ governing the electron cooling in photoexcited bismuth. $G_0$ strongly increases as
a function of electron temperature, which can be traced back to the semi-metallic nature of bismuth.
We also used a thermodynamical model to compute the time evolution of both electron and
lattice temperatures following laser excitation.
Thereby, we simulated the time evolution of (1 -1 0), (-2 1 1) 
and (2 -2 0) Bragg peak intensities measured by Sciaini {\it{ et al}}~ [Nature {\bf 458}, 56 (2009)] 
in femtosecond electron diffraction experiments. The effect of the electron temperature on the Debye-Waller factors 
through the softening of all optical modes across the whole Brillouin zone turns out to be crucial to 
reproduce the time evolution of these Bragg peak intensities.
\end{abstract}
\pacs{63.20.K-, 61.80.Ba, 78.47J-}
\maketitle

Femtosecond time resolved X-ray or electron diffraction emerges as a unique tool to study
laser-induced structural dynamics with a picometer spatial and femtosecond temporal resolution.
A large class of systems including organic compounds\cite{collet_2003}, 
correlated materials\cite{cavalleri_2001, baum_2007}
and semi-metallic systems like graphite\cite{carbone_2008, raman_2008} 
or bismuth\cite{sokolowski_2003, fritz_2007, beaud_2007, johnson_2008, johnson_2009, sciaini_2009} 
have been investigated, providing enlightening information about atomic motions following laser excitation. 
However, it is often difficult to interpret the time evolution of all Bragg peak intensities and to understand
the different time scales showing up in these experiments because models or calculations describing 
laser-induced electron-phonon processes are lacking. 

A first step towards a better understanding of time resolved
X-ray diffraction experiments performed on bismuth\cite{fritz_2007}
was recently achieved\cite{giret_2011}. Indeed, a thermodynamical model based on the
assumption that an electron temperature $T_e$ and a lattice temperature $T_l$ 
can be defined at any time delay following the 
laser pulse arrival, was shown to reproduce the (111) bragg peak intensities measured by 
Fritz {\it{et al}}\cite{fritz_2007} for time delays up to $\sim 2.5$ ps. 
While most of the parameters encompassed in the model were obtained from
first principles calculations, the effective electron-phonon coupling constant $G_0$ governing the electron cooling
was taken as a temperature independent parameter fitted to reproduce the measured Bragg intensities.
Obviously, knowledge of the temperature dependence of $G_0$ is mandatory
to perform simulations over longer time scales.


In this letter, we computed the phonon spectrum of Bi for a large range of electron temperatures
and calculated the effective electron-phonon coupling constant $G_0(T_e)$ using the theory
developped by Allen\cite{allen_1987} which is applied for the first time to
a material showing a strongly varying electron density of states at the Fermi level. We found that
all the optical modes are affected by an increase of $T_e$ and that $G_0$ strongly increases as $T_e$
increases. Then, we solved the model proposed in Ref.\cite{giret_2011} 
and simulated the intensity decay 
of Bragg peaks measured by
Sciaini {\it{ et al}}\cite{sciaini_2009} in femtosecond electron diffraction (FED) experiments for time delays
up to 14 ps. Our calculations firmly establish that the softening of all optical modes is crucial to
achieve a good agreement between experiment and theory. This finding parallels that of Johnson
{\it{et al}}\cite{johnson_2009} who scale all phonon modes by an empirical constant factor smaller than one
to reproduce their normalized diffracted intensities.

All the ab-initio calculations were performed within the framework of the local density approximation (LDA) 
using the ABINIT code\cite{gonze_2009}. Spin-orbit coupling was included and an energy cut-off of
15 Hartree in the planewave expansion of wavefunctions as well as a $16\times 16\times 16$ kpoint grid for the
Brillouin zone integration were used. 
We computed the evolution of the phonon spectrum as a function of $T_e$ using density functional
perturbation theory (DFPT)\cite{gonze_1997}.  Our calculations
show that the phonon spectrum is not affected by $T_e$ below 600 K. However, an 
increase of $T_e$ above this critical temperature leads to a redshift of all optical phonon modes.
Figure 1(a) shows the calculated phonon density of states for $T_e=0$ K 
and $T_e=2100$ K. All the optical modes are softened for $T_e=2100$ K whereas the acoustic modes 
are practically unaffected. This effect has already 
been highlighted in other theoretical works\cite{murray_2007, zijlstra_2010} and should be contrasted
with the bond hardening predicted theoretically\cite{recoules_2006} and observed experimentally 
in gold\cite{ernstorfer_2009}.

Knowledge of the phonon spectrum is essential to compute the decay of electron energy
due to electron-phonon interaction. A key dimensionless quantity is the so called generalized Eliashberg
function\cite{allen_1987} defined by
\begin{eqnarray}\label{spectral_function1}
\alpha^2F(\epsilon, \epsilon^\prime, \omega)&=&\frac{2}{\hbar N_k N_q g(\epsilon_F)}
\sum_{{\bf{k}}, {\bf{q}}, n, m, \lambda} |g_{{\bf q}\lambda}^{{\bf{k}} n m}|^2
\delta[\omega-\omega_{\bf{q}\lambda}] \nonumber \\
& &\times \delta[\epsilon-\epsilon_{{\bf{k}}n}]
\delta[\epsilon^\prime-\epsilon_{{\bf{k+q}}m}],
\end{eqnarray}
where the sum is performed on $N_k$ ($N_q$) electron (phonon) wavevectors over the Brillouin zone,
$g(\epsilon_F)$ is the density of states per unit cell at the Fermi level, $\epsilon_{{\bf{k}}n}$
are Kohn-Sham energies for band $n$ and wavevector ${\bf{k}}$ and where the electron-phonon matrix elements read
\begin{equation}\label{matrix_elements}
g_{{\bf q}\lambda}^{{\bf{k}} n m}=\sum_{p,\alpha} \sqrt{\frac{\hbar}{2 M \omega_{\bf{q}\lambda}}}
e_p^\alpha({\bf q}, \lambda) \langle {\bf{k+q}}m |\frac{\delta V^{SCF}}{\delta u_p^\alpha({\bf q})} |{\bf{k}}n\rangle
\end{equation}
where $e_p^\alpha({\bf q}, \lambda)$ is the displacement of the p$^{th}$ atom of mass $M$ in the
direction $\alpha$ for the mode $({\bf q}, \lambda)$ of frequency $\omega_{\bf{q}\lambda}$. By solving
the Boltzmann equation for both the electron and phonon systems, one can show that the variation
of electron energy per unit volume is given by
\begin{equation}\label{electronic_energy1}
\frac{\partial E_e}{\partial t}=G_0(T_e)\times(T_l-T_e)
\end{equation}
 provided that the lattice temperature $T_l$ is larger than the Debye temperature $\theta_D\simeq 119$ K of Bi.
The effective electron-phonon coupling constant $G_0(T_e)$ reads\cite{allen_1987}
\begin{eqnarray}\label{electronic_energy2}
G_0(T_e)& = &\frac{2\pi g(\epsilon_F) k_B}{v} \int d\omega~
\int d\epsilon~ \alpha^2F(\epsilon, \epsilon+\hbar\omega, \omega) \nonumber \\
& & \times [f(\epsilon)-f(\epsilon+\hbar\omega)],
\end{eqnarray}
where $v$ is the unit cell volume, $k_B$ is the Boltzmann constant, 
and $f(\epsilon)$ is the Fermi-Dirac distribution at temperature $T_e$. 
This expression shows that the integration over $\epsilon$ should be done in a range of energy around $\epsilon_F$
whose width increases with $T_e$. The usual approximation is tantamount to neglecting the energy dependence of
$\alpha^2F(\epsilon, \epsilon+\hbar\omega, \omega)$ which is then approximated by 
$\alpha^2F(\epsilon_F, \epsilon_F, \omega)\equiv \alpha^2F(\omega)$ when the phonon energies are also neglected. Thus,
the electron-phonon coupling constant becomes independent of $T_e$ and is given by
\begin{equation}\label{electronic_energy3}
G_0= \frac{\pi}{v\hbar} k_B \lambda\langle \omega^2\rangle  g(\epsilon_F),
\end{equation}
where $\lambda\langle \omega^2\rangle$ is the second moment of the Eliashberg function $\alpha^2F(\omega)$. 
Eq. \ref{electronic_energy3} is justified for low electron temperatures or for all electron temperatures
when the electron density of states (DOS) is only weakly energy dependent. The last assumption is far
from being satisfied since Bi is a semi-metallic material with a strongly varying DOS near the Fermi level. 
Therefore, a temperature dependent $G_0$ is expected. 

\begin{figure}
\vskip1.0truecm
\includegraphics[width=8.5cm]{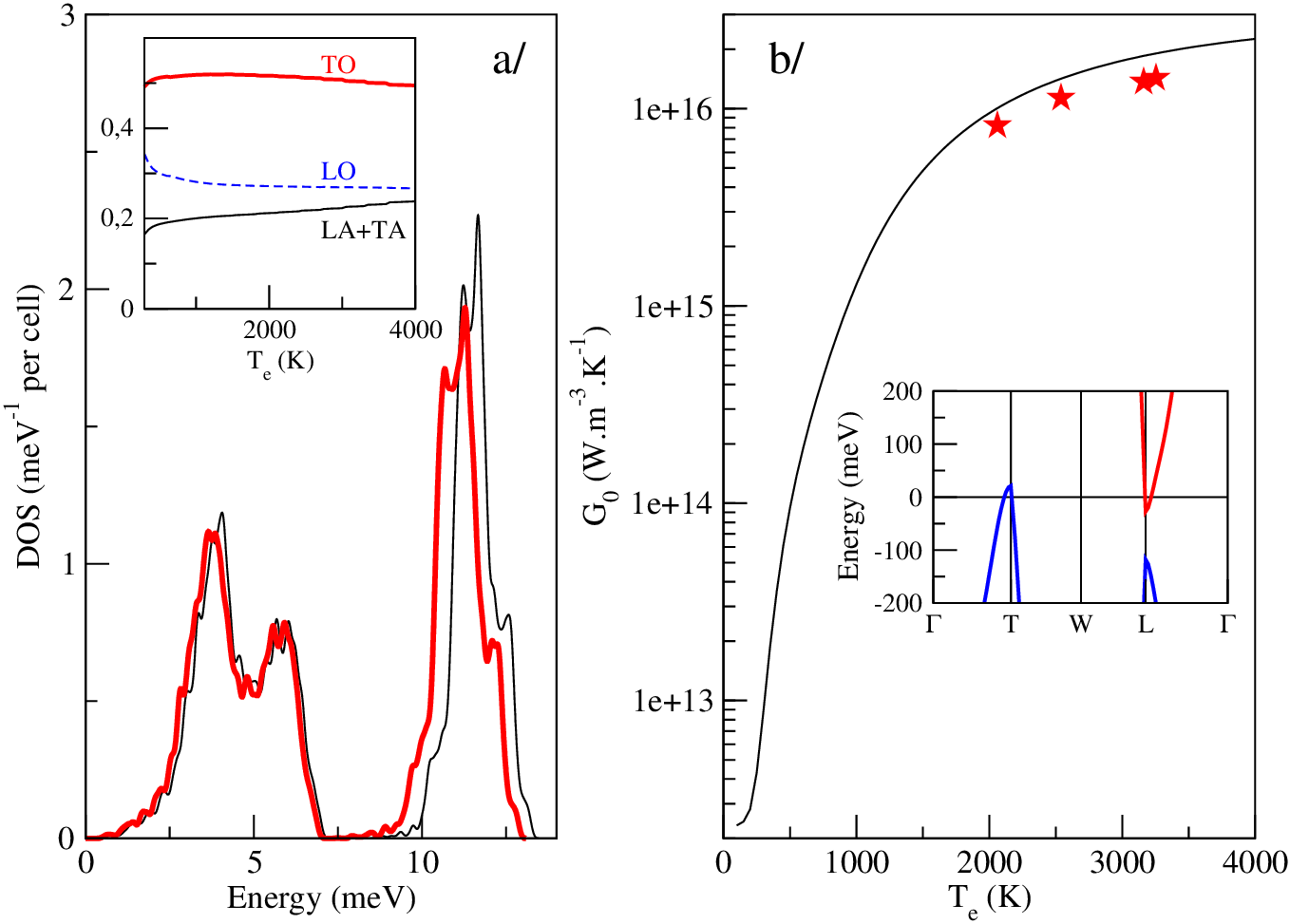}	
\caption{\label{phonon-curves}
(color online) a/ Calculated phonon density of states in meV$^{-1}$ per cell for $T_e=0$ K (thin line)
and $T_e=2100$ K (thick line). The inset shows the fraction of acoustic (TA+LA) and optical (TO and LO)
phonons contributing to the effective electron-phonon coupling $G_0$ for electron temperature 
$T_e$ ranging from 294 K to 4000 K. 
b/ Calculated effective electron-phonon coupling constant $G_0$ (in W.m$^{-3}$.K$^{-1}$) as a function of
$T_e$ (in K) compared to the values fitted in Ref.\cite{giret_2011} to reproduce the time
evolution of (111) Bragg peak intensities measured by Fritz {\it et al}\cite{fritz_2007}. The inset shows
the calculated bandstructure of Bi along high symmetry lines for $T_e=0$ K.
}
\end{figure}

The ab-initio calculated electron-phonon coupling constant $G_0$ 
is shown in Fig. \ref{phonon-curves}(b).
The low temperature value is well described by Eq. \ref{electronic_energy3}
which explains why $G_0$ is so small since $g(\epsilon_F)\sim 1.94\times 10^{-2}$ eV$^{-1}$ per cell.
The huge increase of $G_0$ as
$T_e$ increases can be easily interpreted by looking at the electronic structure
of Bi depicted in the inset of Fig. \ref{phonon-curves}(b). 
At low $T_e$, only intraband scattering processes involving nearly zone center
phonons or interband scattering processes involving phonons connecting the L pocket to the T pockets participate
to the energy exchange between the electrons and the lattice. Thus, the energy is transferred
to only a few phonon modes in this regime. 
When $T_e$ increases, the number of phonon modes participating in the intraband
scattering processes increases tremendously because of larger ${\bf{q}}$-wavectors allowed by momentum 
and energy conservation. Consequently, $G_0$ increases
by four orders of magnitude when $T_e$ increases from 100 K to 4000 K.
The inset in Fig. \ref{phonon-curves}(a)
shows that both optical and acoustic phonons contribute to electron cooling. At 2100 K, transverse optical
phonons (TO), longitudinal optical phonons (LO) and acoustic phonons (TA+LA) respectively contribute
for 51.6, 27.1 and 21.3 \% to the total electron-phonon coupling constant $G_0$. 
In Ref.\cite{giret_2011}, the electron-phonon coupling constants were fitted to reproduce
the time evolution of (111) Bragg peak intensities measured by Fritz {\it{et al}}\cite{fritz_2007} for four
different pump laser fluences. All the fitted values denoted as stars in Fig. \ref{phonon-curves}(b) are
underestimated by $\sim$ 20\% with respect to the calculated values.
However, the (111) Bragg peak intensities are still perfectly reproduced by using our calculated $G_0(T_e)$,
which was also succesfully used to reproduce the time evolution of the electron temperature inferred from
time-resolved photoemission experiments\cite{papalazarou_2012}.
 
\begin{figure}
\vskip1.0truecm
\includegraphics[width=8.5cm]{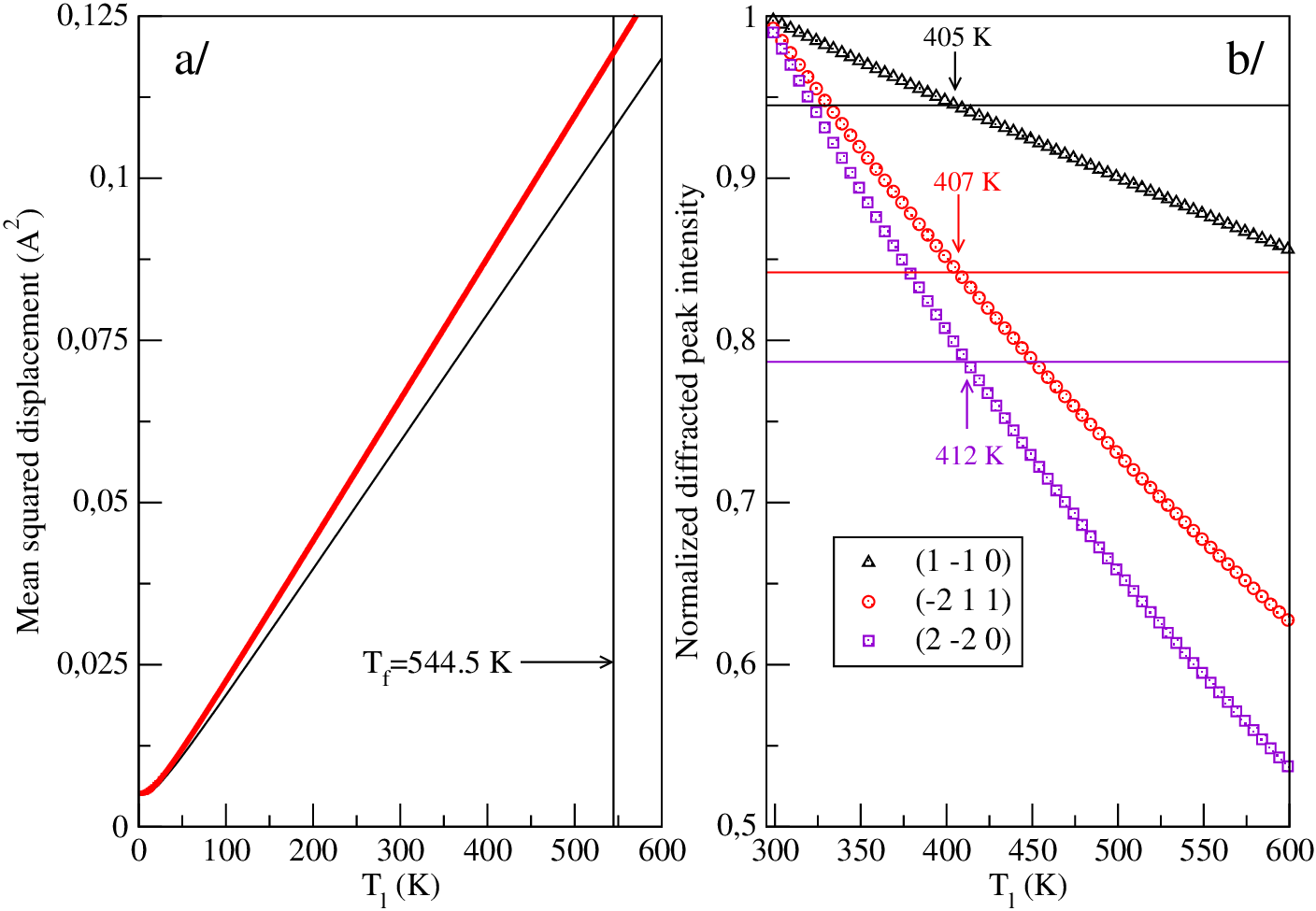}	
\caption{\label{mean-squared-displacment}
(color online) a/Mean square displacements (in \AA$^2$) of one Bi atom as a function 
of lattice temperature T$_l$ (in K)
from {\it ab initio} calculations for $T_e$=294 K and $T_e$=2100 K 
(thin and thick solid lines). 
The vertical line indicates the melting 
temperature of Bi.
b/Calculated normalized intensities defined by 
$\textrm{I}_{\textrm{hkl}}(T_l, T_e=T_l)/
\textrm{I}_{\textrm{hkl}}(T_l=294 ~{\textrm K}, T_e=294 ~{\textrm K})$ 
for (1 -1 0), (-2 1 1) and (2 -2 0)
Bragg peaks as a function of lattice temperature $T_l$ in K. The horizontal lines
are the experimental intensities obtained in FED experiments\cite{sciaini_2009}
 14 ps after the arrival of the 200-fs pump pulse of fluence 0.84 mJ.cm$^{-2}$.
The experimental data are shown in Fig. \ref{time-evolution}(c).
}

\end{figure}

We next calculate the mean square displacement of atoms with respect to their equilibrium 
positions as a function of both lattice temperature $T_l$ and electron temperature $T_e$.
Defining the displacement of the p$^{th}$ atom in the unit cell in the direction $\alpha$
as $u_p^\alpha$, one can show that
\begin{equation}\label{mean_squared_displacement}
\langle u_p^\alpha u_p^\beta\rangle= \frac{\hbar}{2 N_q M }\sum_{{\bf q}, \lambda}
\frac{1}{\omega_{{\bf q}\lambda}} e_p^\alpha({\bf q}, \lambda) e_p^\beta(-{\bf q}, \lambda)
\left[1+ 2 n_{{\bf q}, \lambda }  \right]
\end{equation}
where $n_{{\bf q}, \lambda } $ is the Bose occupation factor at temperature $T_l$ for a
phonon with frequency $\omega_{{\bf q}\lambda}$.
The mean values
defined by Eq. \ref{mean_squared_displacement} have been calculated using the method introduced 
by Lee and Gonze\cite{lee_1995}. Our calculations show that the anisotropy in the mean square
displacements is negligible. Figure \ref{mean-squared-displacment}(a) displays the mean square
displacement of one Bismuth atom (in \AA$^2$) for $T_e=$ 294 K as a function of $T_l$ (thin solid line).
The root mean square displacement increases from 0.070 \AA~ at $T_l$=0K to 0.241 \AA~ at
$T_l$=294 K. The room temperature value is slightly overestimated with respect to the 
experimental value of 0.21 \AA~ extracted from LEED experiments\cite{monig_2005}.
The softening of the optical phonon modes over the whole Brillouin zone 
sketched in Fig.\ref{phonon-curves}(a) has a strong impact on
the mean square displacements which are as higher as the phonon frequencies are lower
(see Eq. \ref{mean_squared_displacement}). Figure \ref{mean-squared-displacment}(a) shows
the increase of the mean square displacements upon increasing $T_e$
from 294 K (thin solid line) to 2100 K (thick solid line) and suggests that the
Bragg peak intensities are affected by $T_e$. 

\begin{figure}
\vskip1.0truecm
\includegraphics[width=8.5cm]{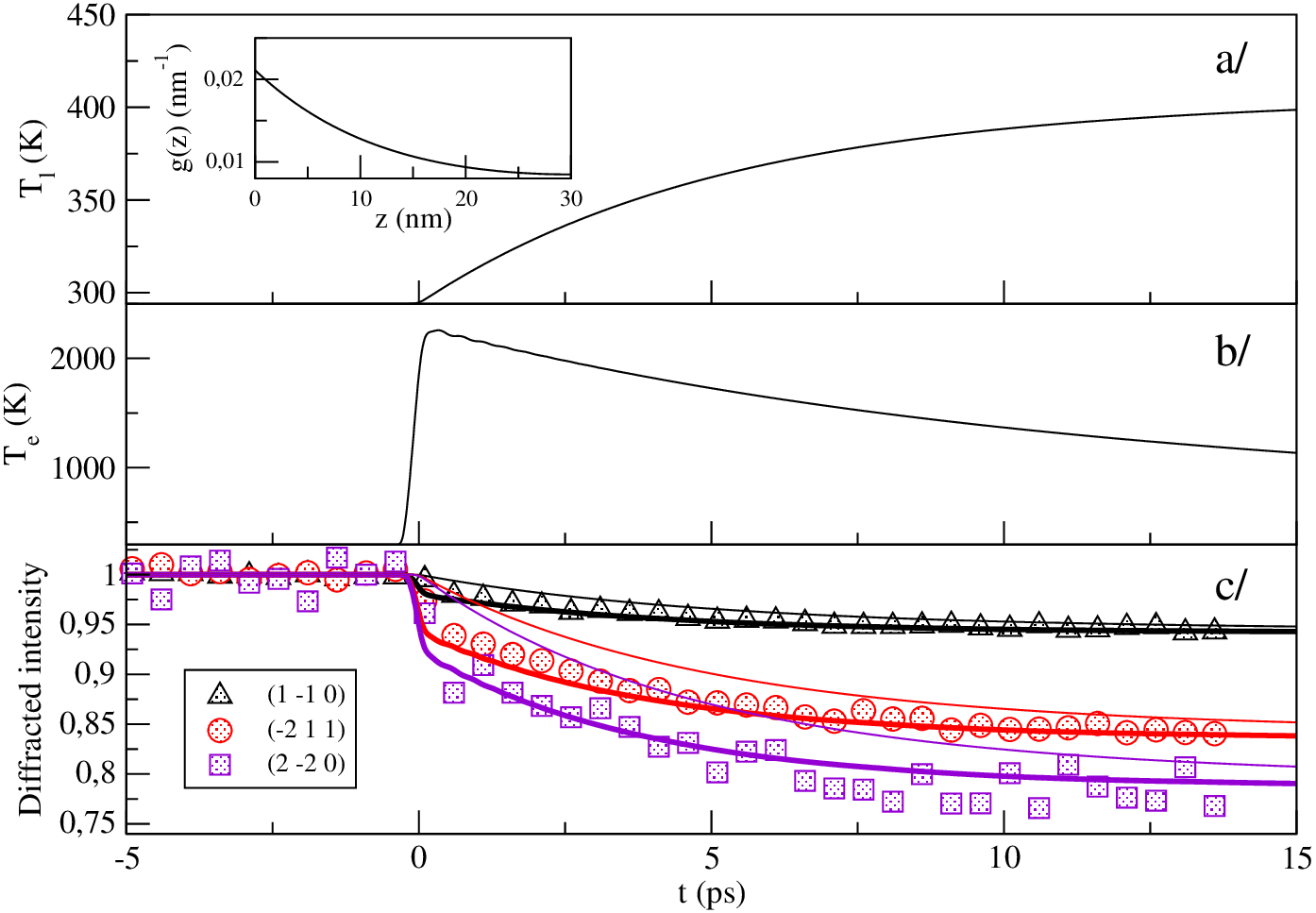}	
\caption{\label{time-evolution}
(color online) a/ and b/ Lattice $T_l$ and electron temperature $T_e$ in K, as a function
of time delay $t$ in ps obtained from the model proposed in Ref.\cite{giret_2011} 
for a 30-nm-thick Bi film excited by a 200-fs laser pulse of incident 
fluence $F_{inc}$=1.11 mJ.cm$^{-2}$ at a wavelength of 775 nm. The inset shows the function $g$ in
nm$^{-1}$ as a function of $z$ in nm.
c/ Normalized intensity decays of (1 -1 0) , (-2 1 1)  
and (2 -2 0)  Bragg peaks measured by Sciaini {\it{ et al}}\cite{sciaini_2009}
as a function of $t$ compared to the theoretical calculations.
The thin lines represent $I_{h,k,l}(T_l, T_e=T_l)/I_{h,k,l}(T_l=294 ~{\textrm K}, T_e=294 ~{\textrm K})$
while the thick lines represent $I_{h,k,l}(T_l, T_e)/I_{h,k,l}(T_l=294 ~{\textrm K}, T_e=294 ~{\textrm K})$. 
}

\end{figure}

The diffracted intensity for a given scattering vector defined by the Miller indices (h,k,l)
is given by
\begin{equation}\label{intensity-definition}
I_{h,k,l}(T_l, T_e)=I_{h,k,l}^0 \exp[-2W(T_l, T_e)]
\end{equation}
where the intensity $I_{h,k,l}^0$, which can be affected by the coherent A$_{1g}$ phonon
coordinate\cite{giret_2011}, is reduced by the so-called Debye-Waller factor.
Here $W(T_l, T_e)$ reads
\begin{equation}\label{def_W}
W(T_l, T_e)=
\frac{1}{2} \sum_{\alpha,\beta} G_\alpha \langle u_p^\alpha u_p^\beta \rangle G_\beta
\end{equation}
where $\langle u_p^\alpha u_p^\beta\rangle$ is obtained from 
equation \ref{mean_squared_displacement}.
Figure \ref{mean-squared-displacment}(b) displays
the calculated ratios $\textrm{I}_{\textrm{hkl}}(T_l, T_e=T_l)/
\textrm{I}_{\textrm{hkl}}(T_l=294 ~{\textrm K}, T_e=294 ~{\textrm K})$
for (1 -1 0), (-2 1 1) and (2 -2 0) Bragg peaks as a function of $T_l$. The
horizontal lines correspond to the normalized intensities obtained by 
Sciaini {\it{ et al}}\cite{sciaini_2009} in their FED studies
14 ps after the arrival of the laser pulse. As can be seen from Fig.\ref{time-evolution}(c), 
the experimental normalized diffracted peak intensities are stationary for this time delay.
Assuming that the lattice is at equilibrium with the electron system for
larger time delays, we can deduce from our calculated normalized intensities three different
lattice temperatures shown by arrows in Fig.\ref{mean-squared-displacment}(b) which are close
to each other. The temperature $T_{l,eq}$ reached by the lattice, when averaging these three values, 
is found to be 408 K and is  smaller than the temperature $T_{l,eq}\simeq 460$ K obtained
by Sciaini {\it{ et al}}\cite{sciaini_2009} on the basis of a 
parametrized Debye-Waller model\cite{gao_1999}. 

In order to simulate the time-resolved FED experiments performed 
by Sciaini {\it{ et al}}\cite{sciaini_2009} on a free standing Bi film of thickness L=30 nm 
excited by a near-infrared laser pulse whose FWHM is $t_w=200$ fs, we used the model proposed
in Ref.\cite{giret_2011} where the source term $P(z,t)$ describing the energy deposited 
per unit volume and time by the laser pulse at time $t$ and at position $z$ reads
\begin{equation}\label{source-term}
P(z,t)=\frac{F_{inc}}{t_w} \sqrt{\frac{4\ln 2}{\pi}} 
g(z) \exp\left[-4 \ln 2\,\frac{t^2}{t_w^2}\right]
\end{equation}
where $F_{inc}$ is the incident fluence and where $g(z)$ is a function plotted in the inset
of Fig.\ref{time-evolution}(a) and calculated by
using both the imaginary and real part of the optical index of Bi for a wavelength 
$\lambda=775$ nm. We note in passing that the absorption coefficient 
$A$ defined by $\int_0^L g(z) dz$ is slightly larger for a 30 nm Bi film ($A=0.36$) 
than for a semi-infinite film ($A=0.31$). The incident fluence $F_{inc}$ needed 
to reproduce the lattice temperature rise extracted from Fig. 
\ref{mean-squared-displacment}(b) is given by $F_{inc}=L\times C_l\times (T_{l,eq}-T_{l,0})/A$
where $T_{l,0}$ is the initial lattice temperature and $C_l$ is the Dulong-Petit
lattice specific heat. Thus, we estimate an incident fluence of 1.11 mJ.cm$^{-2}$. 
This value is overestimated by 27 \% with respect to the measured incident fluence 
of 0.84 mJ.cm$^{-2}$, which might be traced back to the problem of making an accurate 
measurement of the laser fluence.

Using our calculated effective electron-phonon coupling constant $G_0$ shown in Fig. 
\ref{phonon-curves}(b), we numerically solved the three coupled differential equations underlying
the model proposed in Ref.\cite{giret_2011}. Thereby, we obtained the spatial and temporal evolution
of the coherent phonon coordinate $u$ and of both electron $T_e$ and lattice temperature $T_l$ following the
arrival of the 200 fs laser pulse on the sample. As shown in Fig. \ref{time-evolution}(b), 
the electron temperature reaches its maximum value $T_{e,max}\sim 2257$ K 
only 0.3 ps after the arrival of the laser pulse. 
After a strong overheating of the electron system, the electron temperature decreases
whereas the lattice temperature increases.
At $t=$ 14 ps, the lattice is not at equilibrium with the electron system due to the slow down of electron
cooling for low electron temperatures. However, the lattice temperature almost reaches $T_{l,eq}\sim 408$ K because
the energy stored in the electron system is very small.


The normalized intensities of (1 -1 0), (-2 1 1) and (2 -2 0) bragg peaks measured by Sciaini {\it et al}
\cite{sciaini_2009} in FED experiments are shown in Fig. \ref{time-evolution}(c). 
These intensities are compared to the theoretical intensities (thick solid lines) 
defined by $I_{h,k,l}(T_l, T_e)/I_{h,k,l}(T_l=294 ~{\textrm K}, T_e=294 ~{\textrm K})$. 
The agreement between theory and experiment is noteworthy given the fact that all parameters of the model
have been obtained from {\it{ab-initio}} calculations.
In order to point out the role played by $T_e$, we have also displayed 
$I_{h,k,l}(T_l, T_e=T_l)/I_{h,k,l}(T_l=294 ~{\textrm K}, T_e=294 ~{\textrm K})$ (thin solid lines) in Fig. \ref{time-evolution}(c).
Our results show that it is not possible to reproduce the normalized intensity decays
 if we assume that the electron system is at equilibrium with the lattice. The Bragg intensities are
systematically overestimated with respect to experiments. From a physical point of view, 
the softening of the optical phonon modes across the whole brillouin zone when 
$T_e$ is larger than $600$ K leads to a strong decrease of the diffracted 
intensities and brings the theoretical calculations in close agreement with the experimental results.

In conclusion, we computed the effective electron-phonon coupling constant using first-principles
density functionnal theory and found that it strongly depends on the electron temperature. In
addition, we reproduced the time evolution of Bragg peak intensities measured by Sciaini {\it{ et al}}\cite{sciaini_2009} in FED experiments and found that the decay of these intensities is not only
due to the increase of the lattice temperature but also to the redshift of all optical modes arising
from an increase of the electron temperature.

Calculations were performed using HPC resources from GENCI-CINES (project 095096).
We aknowledge G. Sciaini and R.J.D. Miller for providing us with their experimental results 
and also for enlightening discussions.


\begin{thebibliography}{}

\bibitem{collet_2003}
{E. Collet {\it et al}}, {Science} {\bf  300}, 612 (2003).

\bibitem{cavalleri_2001}
{A. Cavalleri {\it et al}}, {Phys. Rev. Lett.} {\bf  87}, 237401 (2001).

\bibitem{baum_2007}
{P. Baum , D. Yang and A. H. Zewail}, {Science} {\bf  318}, 788 (2007).

\bibitem{carbone_2008}
{F. Carbone, P. Baum, P. Rudolf and A. H. Zewail}, {Phys. Rev. Lett.} {\bf  100}, 035501 (2008).

\bibitem{raman_2008}
{R. K. Raman {\it et al}}, {Phys. Rev. Lett.} {\bf  101}, 077401 (2008).

\bibitem{sokolowski_2003}
{K. Sokolowski-Tinten {\it et al}}, {Nature} {\bf  422}, 287 (2003).

\bibitem{fritz_2007}
{D.M. Fritz {\it et al}},
{Science} {\bf  315}, 633 (2007).

\bibitem{beaud_2007}
{P. Beaud {\it et al}}, {Phys. Rev. Lett.} {\bf  99}, 174801 (2007).

\bibitem{johnson_2008}
{S. L. Johnson {\it et al}}, {Phys. Rev. Lett.} {\bf  100}, 155501 (2008).

\bibitem{johnson_2009}
{S. L. Johnson {\it et al}}, {Phys. Rev. Lett.} {\bf  102}, 175503 (2009).

\bibitem{sciaini_2009}
{G. Sciaini {\it et al}}, Nature {\bf 458}, 56 (2009).

\bibitem{giret_2011}
{Y. Giret, A. Gell\'e and B. Arnaud}, {Phys. Rev. Lett.} {\bf  106}, 155503 (2011).

\bibitem{allen_1987}
{P. B. Allen}, {Phys. Rev. Lett.} {\bf  59}, 1460 (1987).





\bibitem{gonze_2009}
{X. Gonze {\it et al}}, {Comput. Phys. Commun.} {\bf  180}, 2582 (2009).




\bibitem{gonze_1997}
{X. Gonze}, 
{Phys. Rev. B} {\bf  55}, 10337 (1997).



\bibitem{murray_2007}
{E.D. Murray, S. Fahy, D. Prendergast, T. Ogitsu, D.M. Fritz, and D.A. Reis}, 
{Phys. Rev. B} {\bf  75}, 184301 (2007).

\bibitem{zijlstra_2010}
{E.S. Zijlstra, L.E. Diaz-Sanchez, and M.E. Garcia}, {Phys. Rev. Lett.} {\bf  104}, 029601 (2010).

\bibitem{recoules_2006}
{V. Recoules, J. Cl\'erouin, G. Z\'erah, P.M. Anglade, and S. Mazevet}, {Phys. Rev. Lett.} {\bf  96}, 055503 (2006).

\bibitem{ernstorfer_2009}
{R. Ernstorfer {\it et al}}, {Science} {\bf  323}, 1033 (2009).



\bibitem{papalazarou_2012}
{E. Papalazarou, {\it et al}}, {Phys. Rev. Lett.} {\bf  108}, 256808 (2012).



\bibitem{lee_1995}
{C. Lee and X. Gonze}, {Phys. Rev. B} {\bf  51}, 8610 (1995).


\bibitem{monig_2005}
{H. M\"onig {\it et al}}, {Phys. Rev. B} {\bf  72}, 085410 (2005).







\bibitem{gao_1999}
{H.X. Gao and L.M. Peng}, {\it{Acta Cryst}}. A{\bf 55}, 926 (1999).





\end{thebibliography}
\end{document}